\documentclass[pre,twocolumn]{revtex4-1}
\usepackage{graphicx}
\usepackage{dcolumn}
\usepackage{bm}
\usepackage{amssymb}
\newcommand{\comment}[1]{}

\begin{document}

\title{Deformation-induced accelerated dynamics in polymer glasses}

\author{Mya Warren}
\email{mya@ctbp.ucsd.edu}
\affiliation{Center for Theoretical Biological Physics, University of California at San Diego}
\author{J{\"o}rg Rottler}
\affiliation{Department of Physics and Astronomy, University of British Columbia}

\date{\today}

\begin{abstract}
Molecular dynamics simulations are used to investigate the effects of deformation on the segmental dynamics in an aging polymer glass. Individual particle trajectories are decomposed into a series of discontinuous hops, from which we obtain the full distribution of relaxation times and displacements under three deformation protocols: step stress (creep), step strain, and constant strain rate deformation. As in experiments, the dynamics can be accelerated by several orders of magnitude during deformation, and the history dependence is entirely erased during yield (mechanical rejuvenation). Aging can be explained as a result of the long tails in the relaxation time distribution of the glass, and similarly, mechanical rejuvenation is understood through the observed narrowing of this distribution during yield. Although the relaxation time distributions under deformation are highly protocol specific, in each case they may be described by a universal acceleration factor that depends only on the strain.
\end{abstract}



\maketitle

\section{Introduction}
\label{sec:introduction} 

Although it has long been recognized that mechanical deformation can
modify the structural relaxation dynamics of polymer glasses,
obtaining a complete picture of the deformation-induced mobility
remains an important challenge. It is most often assumed that
mechanical stress lowers the activation barriers to molecular
mobility, thereby allowing the solid to yield and flow. This simple
idea, first proposed by Eyring in 1936 \cite{Eyring}, remains the
central tenet of several models of plasticity in amorphous solids
\cite{Tervoort1996,Schweizer_PRL98}. Other control variables have also
been proposed to explain the increased molecular mobility under
deformation. In the Soft Glassy Rheology (SGR) model, activated
barrier crossings are accelerated by the local strain on mesoscopic
regions of the glass \cite{Sollich_PRL78,Sollich_JR44}. In the Shear
Transformation Zone (STZ) theory, deformation is caused by a
population of bistable states (shear transformation zones) that can
flip from one orientation to another in response to stress
\cite{Falk_PRE57,Langer_SM54}. The stress induced transition between the bistable states is modeled using a variant of the Eyring
model, and the population of shear transformation zones is controlled
by an effective temperature which explicitly depends on the strain
rate $\dot{\epsilon}$ \cite{Langer_PRE70,Langer_PRE76}. Still other models have
proposed the free volume \cite{Struik}, and the configurational
entropy \cite{Shay1996} as the important variables to describe the
deformation-induced mobility. There is clearly no consensus on the
correct way to think about deformation in glasses.

Experimentally, the structural relaxation times of deformed polymer
glasses have traditionally been studied through rheological measurements,
where the segmental dynamics are inferred from the mechanical response
\cite{Struik, Gacougnolle_Poly43, McKenna_O'Connell2002,
McKenna_Zapas1986, McKenna_Waldron1995}. In recent years, experiments
and simulations have allowed researchers to directly probe the
microscopic segmental dynamics during deformation. Solid-state NMR
experiments examined the dynamics of glassy Nylon-6 near $T_g$ under
uniaxial extension and found significant deformation-induced mobility
near yield \cite{Gleason2000}. Lee et al.~performed a series of
experiments investigating the segmental relaxation dynamics in solid
poly(methyl methacrylate) through an optical photobleaching technique
that probes the reorientation of small dye molecules embedded in the
polymer matrix during a step stress (creep) experiment
\cite{Ediger_JCP128,Ediger_Science323}.  Relaxation times were found
to be significantly accelerated compared to the undeformed sample, and
in fact the segmental mobility was close to linearly
correlated with the strain rate. During plastic flow, the distribution
of relaxation times was also narrowed, which was interpreted as
evidence that the dynamics has become more homogeneous.

Molecular dynamics simulations have confirmed these results through
direct measurement of the segmental dynamics. Lyulin et al.~performed
simulations on united atom models for polystyrene and polycarbonate
during constant strain rate deformation \cite{Lyulin_EPL71}. The
segmental mobility was monitored through the monomer mean-squared
displacement and increased dramatically beyond the yield
point. Similarly, simulations of polyethylene under compressive strain
showed that the dihedral transition rate in between trans/gauche
states was increased due to the deformation even in the sub-yield
regime \cite{Boyce_PRL89}. Several simulation studies have also
confirmed the strong correlation between the relaxation time and
strain rate that has been seen in experiment. Riggleman et
al.~simulated a bead-spring polymer glass during non-linear creep
\cite{Ediger_PRL99,Riggleman_MM41}. The covalent bond autocorrelation
and intermediate scattering functions both decayed more rapidly in the
deformed sample, and it was found that the segmental mobility was
strongly correlated to the strain rate under both tensile and
compressive loading.

For polymer glasses in particular, the effects of physical aging
further complicate the analysis of the dynamics under deformation, as
the mechanical response depends sensitively on the history of the
glass. Non-equilibrium relaxations cause relaxation times to grow, and
hence the glass becomes stiffer with the time since it was
formed. Since the age of a sample is often characterized by its
relaxation times, the deformation-induced acceleration of the dynamics
has been associated with ``mechanical rejuvenation'' of aging glasses
\cite{Struik}. Indeed, in the post-yield flow regime, all history
dependence is erased. Hasan and Boyce investigated this phenomenon
using differential scanning calorimetry on polymer glasses during
constant strain rate deformation \cite{Boyce_Polymer34}. While, at
sub-yield strains, the aged glasses exhibited different mechanical
properties and enthalpies, these differences were found to disappear
once the glasses began to flow. Simulations have provided further
insight into the nature of rejuvenation. Utz et al.~found that the
history dependence of the inherent structure energies and the pair
correlation functions of the binary metallic Lennard-Jones glass
(BMLJ) were also erased during yielding \cite{Utz_PRL84}. Post-yield,
these parameters took values more typical of rapidly quenched, or
unaged samples. Further analysis, however, seems to indicate that the
rejuvenated glass is not truly the same as a younger glass
\cite{Lacks_PRL93,Lyulin_PRL99}. 

Below yield, the claim of mechanical rejuvenation is even more
controversial \cite{McKenna_JPhys15}. In this regime, deformation
increases molecular mobility at the same time as aging decreases it
\cite{Lee_PhD2009}, and experiments show that both the volume
\cite{McKenna_Poly31} and the relaxation times \cite{Lee_PhD2009} of
the glass quickly return to their previous states once the load is
removed \cite{Warren_PRE78}. In this report, we
reserve the term ``rejuvenation'' for the erasure of aging due to
deformation, without comment on whether the yielded state truly
resembles a younger glass or is something else altogether.

Mobility and aging in previous studies has been measured using
spatially averaged correlation and response functions; however, it is
becoming increasingly clear that glasses are both dynamically and
mechanically heterogeneous. A number of simulations have studied the
spatial heterogeneity of structural relaxations under deformation in
the limit of low temperatures and small shear rates
\cite{Maloney_PRL93,Barrat_EPJ20}. Results show that the elementary
relaxations responsible for plastic deformation are strongly
localized, involving a region of several diameters across. They are
also intermittent: rapid relaxation events are separated by periods of
perfectly affine deformation. Furthermore, the stress and elastic
modulus are also locally heterogeneous \cite{Barrat_PRE80,
dePablo_PRL93} and structural relaxations occur predominantly in weak
regions of the glass \cite{Barrat_PRE80}. 

We use molecular dynamics simulations of an aging polymer glass to
investigate the complex interaction between aging and deformation at
the microscopic level, and to identify the deformation variables
important to the description of the deformation-induced
mobility. Three deformation protocols are employed: a step stress
(creep), a constant strain rate deformation, and a step strain (stress
relaxation). The microscopic relaxation events are identified as
discontinuous ``hops'' in the individual monomer trajectories which
occur in between periods of affine deformation, and the full
distribution of hop times and displacements are evaluated. 
Previously, we reported that the changes to these distributions in
three different deformation protocols and for different ages exhibit a
universal dependence on global strain \cite{Warren_PRL104}. 
In the present paper, we provide a more
detailed account of these findings, present additional data, and
expand on the analysis.  In Sections \ref{sec:sims} and
\ref{sec:hopping}, we outline the details of our simulations and our
methodology for detecting hops. Section \ref{sec:results} presents the
results of these simulations, and in Section \ref{sec:conclusions} we
discuss our finding and conclusions.  We also present in the Appendix
further details of the hop analysis technique used to determine the
relaxation time distributions.

\section{Simulations}
\label{sec:sims}
We perform molecular dynamics (MD) simulations of a well-known
bead-spring polymer model for polymer glasses
\cite{Bennemann_PRE57,Baschnagel_JPCM12}.  The beads interact via a
non-specific van der Waals interaction given by a 6-12 Lennard-Jones
potential, and the covalent bonds are modeled with a stiff spring that
prevents chain crossing \cite{Kremer_Grest}.  The Lennard-Jones
interaction is truncated at $1.5$ times the bead diameter and adjusted
vertically for continuity. Unless otherwise noted, all polymers have a
length of 10 beads, and each sample contains two thousand polymers. To
check that results are insensitive to chain length, we also examined
selected cases with chains of length 100 (slightly longer than the
entanglement length \cite{Kremer2004}). The simulation box is
initially cubic with periodic boundary conditions. All results will be
given in units of the diameter $a$ of the bead, the mass $m$, and the
Lennard-Jones energy scale, $u_0$. The reference time scale is
$\tau_{LJ} = \sqrt{ma^2/u_0 }$.

To create a glass, the ensemble of polymers is first relaxed at
constant volume and a melt temperature of $1.2$ and then cooled to
form a solid at $T = 0.25$ ($T_g \approx 0.35$
\cite{Rottler_PRE68}). The density of the melt is chosen such that
after cooling the pressure on the box is zero. The glass is then aged
for various wait times $t_w$ at constant temperature and zero pressure
using a Nos\'e-Hoover thermostat/barostat. For the creep protocol, a
constant uniaxial tensile stress in the x-direction is ramped up over a
time of $75$ and then held constant using the Nos\'e-Hoover
barostat. Four different stresses are applied $\sigma=0.2$, 0.3,
0.4, and 0.5 in order to investigate a range of behaviour from the
linear regime to yielding and flow. In the second protocol, a constant
shear rate is applied by rescaling the particle positions and the
simulation box in the $x$ direction. Three strain rates are used:
$\dot{\epsilon}=2.2\times 10^{-7}$, $8.9\times 10^{-7}$, and
$8.9\times 10^{-6}$. For the final protocol, a step strain is applied
over a time of $75$ and then held constant. Strains of 0.01 to
0.04 are evaluated. In all three cases, the stress is maintained at
zero in the other two directions, and their dimensions respond
dynamically to the imposed deformation through the Nos\'e-Hoover
barostat.

\section{Hopping dynamics}
\label{sec:hopping}

\begin{figure}[tbp]  
\begin{center}
\includegraphics[width=8cm]{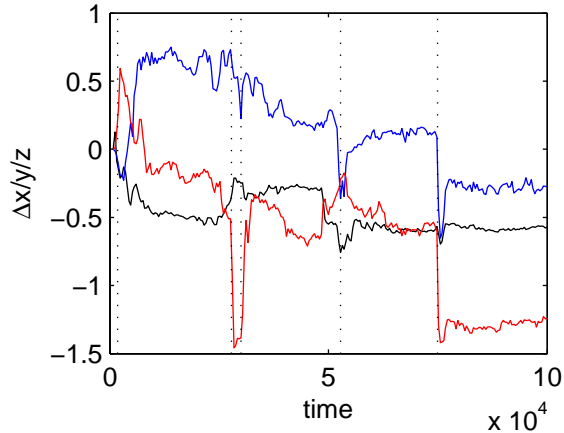}
\caption{Non-affine displacements $\Delta x$, $\Delta y$, and
$\Delta z$ (black/blue/red) for a monomer undergoing creep at $\sigma
= 0.5$. The vertical dotted lines indicate relaxations identified by
our hop detection algorithm. The total strain during this time is
approximately 0.8.}
\label{fig:traj}
\end{center} 
\end{figure}

In addition to monitoring the mechanical response in the deformation
protocols described above, we are interested in understanding more
directly how the deformation accelerates the particle relaxation
dynamics. Structural relaxations are identified on the scale of a
single particle or monomer segment by monitoring a subset of five
thousand particle trajectories chosen randomly from the simulation
volume. Figure \ref{fig:traj} shows an example of the non-affine
particle displacement during a creep experiment (the portion of the
displacement due to the macroscopic deformation of the simulation
volume has been removed). The trajectories take the form of long
periods of vibration where their motion is predominantly affine,
punctuated by discrete hopping events.

Single particle hopping dynamics have been studied in simulations of
metallic liquids and glasses \cite{Vollmayr_JCP121,Vollmayr_PRE72}, in
kinetically constrained models of supercooled liquids
\cite{Chandler_JCP127}, and in experiments of colloidal glasses
\cite{Chaudhuri_JPCM} and granular materials \cite{Dauchot_PRL102},
and have been very useful in understanding the nature of heterogeneous
dynamics in soft glassy materials. To detect the individual hops, a
running average and standard deviation is computed for the non-affine
displacement of each particle over a time of $400$. The usual
method of defining a hop is through a threshold in the displacement of
the mean particle position, where the threshold is determined in
relation to the vibrational amplitude \cite{Vollmayr_JCP121,
Chandler_JCP127, Chaudhuri_JPCM, Dauchot_PRL102}. However, we have
observed that frequently particles experience large, intermittent
fluctuations in position with little net displacement. These events
can not be disregarded as hops, as the displacement is often much larger than the cage size. The ambiguity in
determining a hop from a single trajectory is related to the fact that
the single particle hop is only a small part of a collective
relaxation event. In this work, we therefore identify hops through a
threshold in the standard deviation of the displacement, rather than
the mean; that is, we define relaxation events by the
\textit{activity} of the particle, rather than the final
state. Further details of the procedure are discussed in the
Appendix. Once the trajectories are reduced to a series of hops, we
construct the probability distribution functions for the first hop
time $p(t_1)$ (the time from the beginning of the measurement to the
first hop), the persistence time $p(\tau)$ (the time in between all
subsequent hops or the residence time for a caged configuration), and
the hop displacements $p(dx)$.

\section{Results}
\label{sec:results}
\subsection{Aging at zero pressure}
\label{sec:aging}

At finite temperature, the undeformed glass continues to evolve in a
non-ergodic manner due to activated thermal processes. The
non-equilibrium relaxation dynamics causes aging in all of the
properties of the glass with the wait time $t_w$ since
vitrification. The volume decreases logarithmically with the wait
time, as does the potential energy. The most striking results of aging
are, however, seen in the dynamical properties of the glass. Glasses
obey ``time-waiting time superposition'': two-time correlation and
response functions can be rescaled by the wait time $C(t,t_w) =
C_{vib}(t) + C_{conf}(t/t_w^{\mu})$, where $\mu$ is the aging
exponent and $t$ is the measurement time. Although glasses are characterized by a wide distribution of relaxation times, it appears as though they are all rescaled
identically during aging.

\begin{figure*}[tbp]  
\begin{center}
\includegraphics[width=17.5cm]{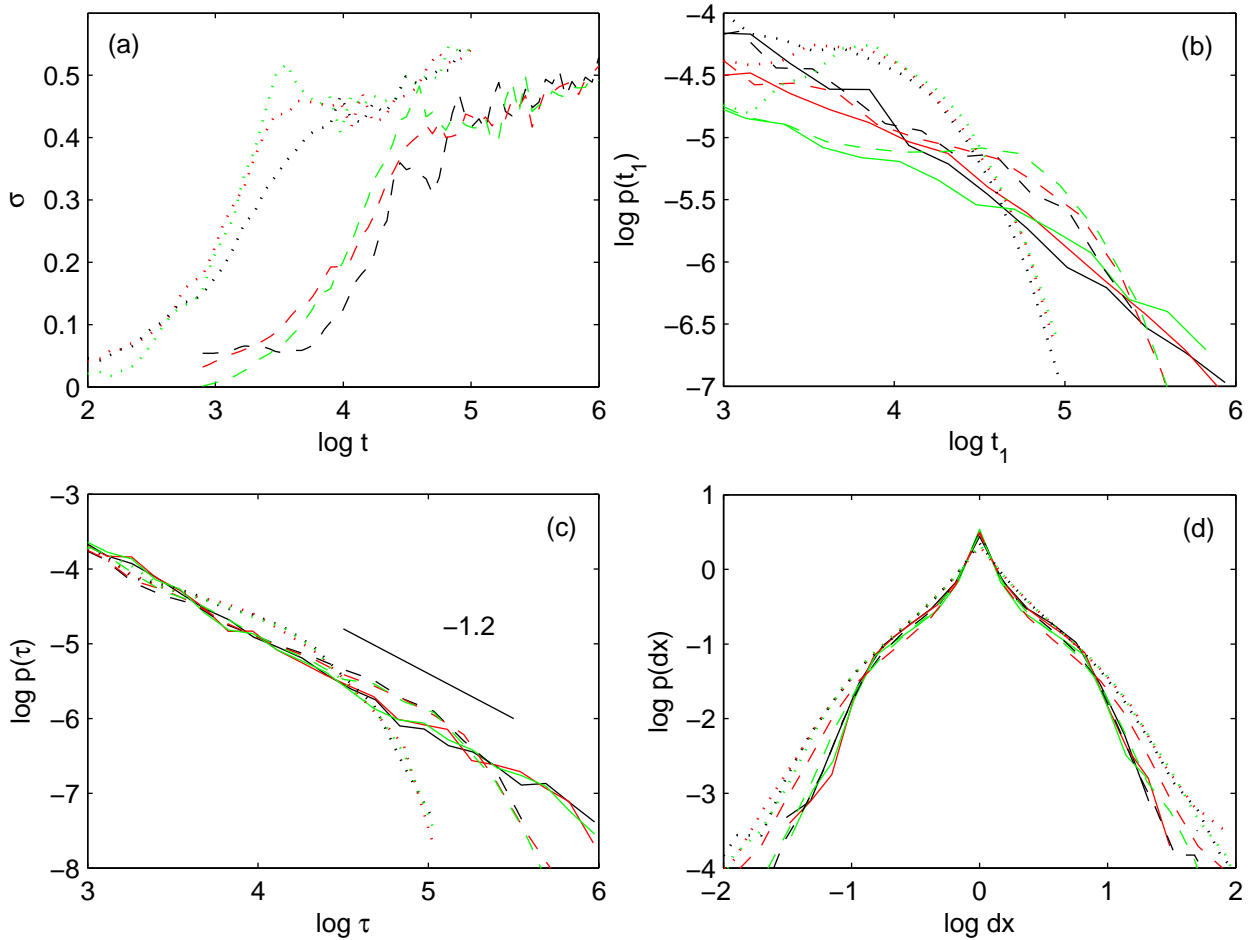}
\caption{Relaxation dynamics during the constant strain rate
experiment. (a) Stress versus time for $\dot{\epsilon} = 8.9\times
10^{-7}$ (dashed) and $\dot{\epsilon} = 8.9\times 10^{-6}$
(dotted). Distributions of (b) first hop times, (c) persistence times,
and (d) displacements for the undeformed sample (solid lines) and
$\dot\epsilon= 8.9\times 10^{-7}$ (dashed) and $8.9\times 10^{-6}$
(dotted). In all plots $t_w = 750$ (black), 7500 (red), and 75000
(green). Straight lines indicate power law with the given slopes.}
\label{fig:strain_rate}
\end{center} 
\end{figure*}

Previously, we reported on the hop statistics in a polymer and binary
metallic Lennard-Jones glass undergoing simple aging after a rapid
quench \cite{Warren_EPL88}. Hop
statistics for the polymer glass can be seen here again as solid lines
in Figs.~\ref{fig:creep}(b-d), and also in
Figs.~\ref{fig:strain_rate}(b-d) and ~\ref{fig:strain}(b-d). The first
hop time distributions (Fig.~\ref{fig:creep}(b)) take the form of two
power laws, and the intersection between the two increases with 
$t_w^\mu$ in the same way as the correlation and response functions. This time seems to be correlated with the cage escape time $\tau_\alpha$ where structural relaxations become important in the glass. Alternatively, Figure \ref{fig:creep}(c) shows that
the distribution of persistence times decays as a single,
age-independent power law, $p(\tau) \sim \tau^{-1.2}$. In ref.~\cite{Warren_EPL88}, we showed that aging can be understood as the
direct result of the wide distribution of persistence times which, in
the thermodynamic limit, has an infinite mean. Evolving the system of particles using a continuous time random walk parameterized with the initial condition $p(t_1,t_w=0)$, which reflects the quench, and the persistence time distribution $p(\tau)$ is sufficient to predict the observed changes to the first hop time distributions $p(t_1,t_w)$ during aging.

A power law form for $p(\tau)$ was also found for the BMLJ supercooled liquid by using a rigorous energy landscape approach to detect transitions \cite{Heuer_PRL91}. Doliwa and Heuer performed frequent energy minimizations in order to detect collective relaxations in between adjacent metabasins in the potential energy landscape. The persistence time distributions were found to take the form of two power laws: at short times the transitions are continuous and liquid-like, and at long times the transitions are dominated by uncorrelated hops between metabasins. The decay at long time becomes less steep with temperature, approaching $\tau^{-2}$ as the glass transition temperature is approached from above; that is, $\langle \tau \rangle$ is finite in the liquid state. Vollmayer-Lee measured the persistence time distributions in the binary Lennard-Jones glass using a single particle hop detection algorithm which differentiates between hops that lead to an entirely new position (irreversible hops) and hops back and forth between two average positions (reversible hops) \cite{Vollmayr_JCP121}. $p(\tau)$ was again found to decay in a weak power law ($\tau^{-0.84}$), independent of history and temperature in the glassy state. 


\begin{figure}[tbp]  
\begin{center}
\includegraphics[width=8cm]{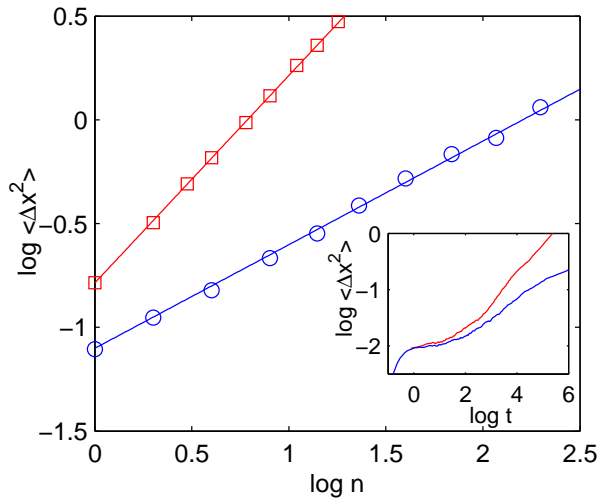}
\caption{Mean-squared displacement as a function of the number of hops $n$
that a particle has experienced for the polymer model ($\circ$) and
the BMLJ model ($\square$) (simulation details of the BMLJ results
can be found in \cite{Warren_EPL88}). Lines have slopes of $1/2$ (blue) and 1
(red). Inset shows the corresponding mean-squared displacement for
both models with time.}
\label{fig:var_vs_n}
\end{center} 
\end{figure}

Figure \ref{fig:creep}(d) shows that the hop
displacement distribution $p(dx)$ also does not depend on the wait
time. $p(dx)$ is sharply peaked around zero and decays rapidly at
large $dx$. The rapid decay of the tails results from the connectivity
of the polymer glass; the tails of $p(dx)$ are purely exponential in
the binary metallic Lennard-Jones (BMLJ) glass \cite{Warren_EPL88}. 
Polymer specific dynamics is also apparent in the diffusion behaviour
of the polymer glass. The mean-squared displacement $\langle \Delta
x(t)^2 \rangle$ of both the polymer model and the binary metallic
Lennard-Jones glass show the two-step relaxation behaviour
characteristic of glassy materials (inset
Fig.~\ref{fig:var_vs_n}). However, in the cage-escape regime, $\langle
\Delta x(t)^2 \rangle$ increases much more rapidly in the metallic
glass than in the polymer glass. The reason for this is evident if the
mean-squared displacement is plotted versus the number of hops $n$ a
particle has experienced rather than the time elapsed $t$, as
shown in Figure \ref{fig:var_vs_n}. The mean-squared displacement of
the polymer beads increases with the number of hops as $\langle
\Delta x(n)^2 \rangle \sim \sqrt{n}$ (Rouse dynamics) rather than
$\sim n$ for simple molecular glass formers due to the constraints of
the covalent bonds. Alternatively, the aging behaviour of the
relaxation time distributions were found to be qualitatively similar
for the polymer model and the BMLJ glass \cite{Warren_EPL88}. This is a
good indication that the results discussed here apply quite generally
to glassy materials, and the polymer-specific aspects of the dynamics
are seen primarily in the displacements rather than the relaxation
times.

\subsection{Step stress}
\label{sec:creep}

\begin{figure*}[tbp]  
\begin{center}
\includegraphics[width=17.5cm]{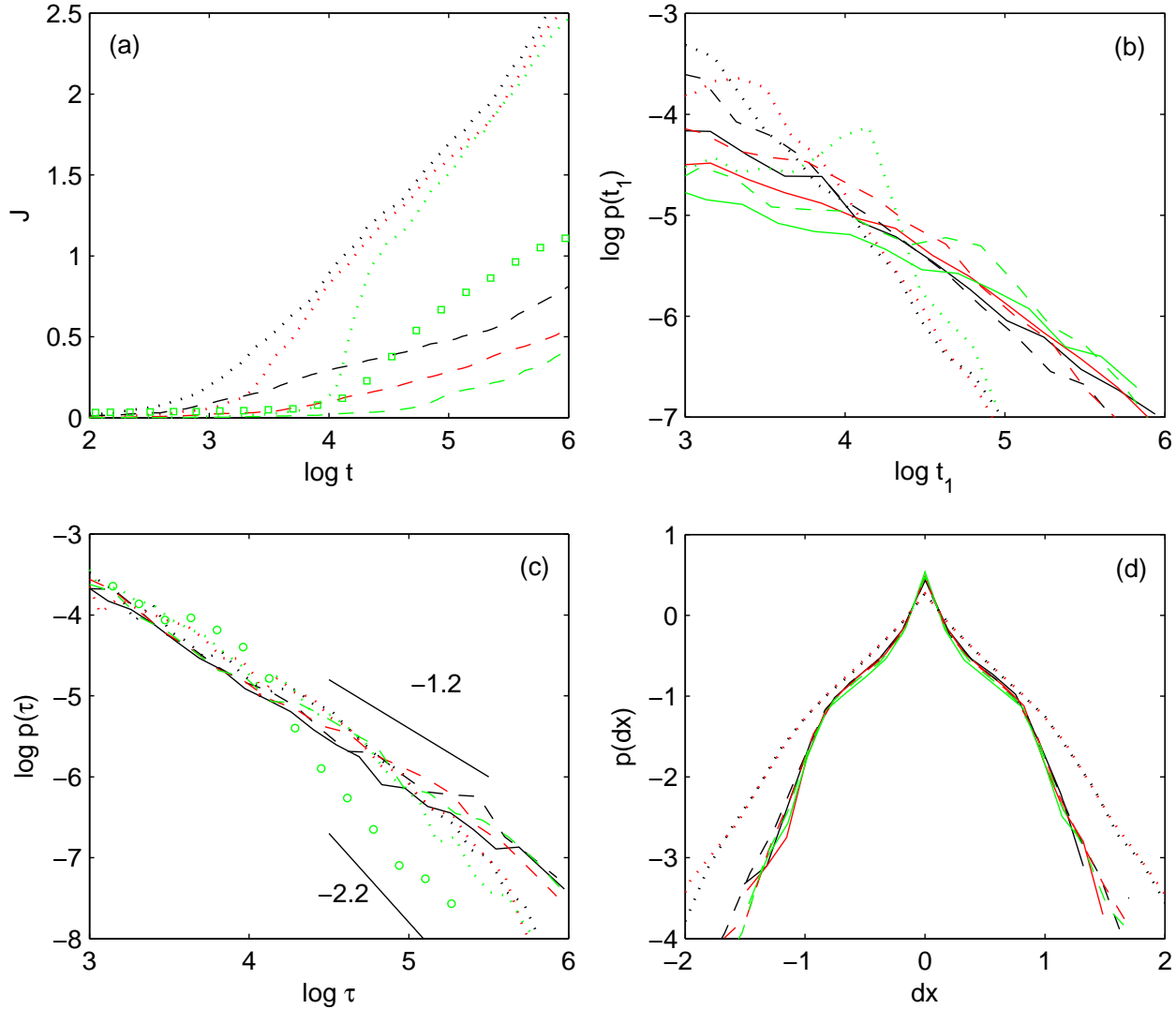}
\caption{Relaxation dynamics during the creep experiment. (a)
Creep compliance for $\sigma = 0.4$ (dashed), 0.5
(dotted). Squares show the $\sigma = 0.5$ curve for chains of length 100. Distributions of (b) first hop times, (c) persistence times,
and (d) displacements for the undeformed sample (solid lines), $\sigma
= 0.4$ (dashed), and $\sigma = 0.5$ (dotted). In all plots $t_w = 750$
(black), 7500 (red), and 75000 (green). Circles in (c) show
$p(\tau)$ for $t<\tau_\alpha \approx 10^4$ (see text) and $t_w = 75000$. Straight
lines indicate power law with the given slopes.}
\label{fig:creep}
\end{center} 
\end{figure*}

We next turn our attention to the effects of deformation on the
relaxation dynamics. Figure \ref{fig:creep} shows the mechanical
response (a), and the hop statistics (b-d) in the step stress
experiment. When a stress $\sigma$ is suddenly applied to the glass,
the strain $\epsilon(t,t_w)$ shows an initial elastic response
followed by slow elongation, or creep, due to structural relaxations
in the glass. In Figure \ref{fig:creep}(a), the creep compliance
$J(t,t_w) = \epsilon(t,t_w)/\sigma$ is plotted as a function of wait
time and stress. At small stress, the creep compliance takes the form
of a stretched exponential at short times, changing to a logarithmic
increase at $t\gg t_w$ due to the effects of aging at long times \cite{Struik,Warren_PRE07}. For increasing
wait time, the glass becomes stiffer and the creep compliance curves
at short times can be superimposed to form a master curve by rescaling
the time variable with a power law in $t_w$,
\begin{equation}
J(t,t_w)=\mathcal{J}(t/t_w^\mu).
\label{eqn:master}
\end{equation}
The aging exponent $\mu$ is approximately $0.8$ at a small stress of $\sigma = 0.3$. 
At $\sigma = 0.5$, Fig.~\ref{fig:creep}(a) shows that the short time
compliance depends on the wait time, but at longer times the glass
yields and all history dependence is lost. The strain rate in this
regime decreases over time due to the effects of strain-hardening.

The hop statistics for the $\sigma = 0.4$ and 0.5 creep experiments
are shown in Fig.~\ref{fig:creep}(b-d), and compared to the
unstressed, aging glass. For $\sigma = 0.5$, the first hop times are
clearly accelerated by the stress. The tails of the distribution decay
with a much steeper power law than the undeformed glass ($t_1^{-2.2}$
vs.~$t_1^{-1.1}$). This corresponds well to experiments, which showed
that in the flow regime, not only are the segmental dynamics
accelerated, but the width of the relaxation time distribution becomes
narrower as well \cite{Ediger_Science323}. This distribution is
narrowed for all stresses tested here, however the changes to $p(t_1)$
are considerably smaller in the sub-yield regime. The persistence time
distributions are also narrowed during the creep experiment. Unlike in
the case of simple aging, $p(\tau)$ during a creep experiment depends
explicitly on both the wait time $t_w$, and the measurement time $t$
since the stress was applied. The distributions shown in Fig.~\ref{fig:creep}(c) are obtained
from the persistence times of cages originating at times much greater
than the cage escape time, where the strain is increasing
logarithmically. The narrowing of
$p(\tau)$ is even more pronounced for cages originating at shorter
times (circles). This result provides an interesting explanation
for mechanical rejuvenation. Since aging can be understood as the
effect of an infinite mean persistence time, then aging ceases when
the tails of $p(\tau)$ decay with an exponent steeper than $-2$, and a
steady state is established which is independent of the age of the
glass. For glasses below yield, the persistence time distribution
remains in the aging regime, whereas at $\sigma = 0.5$, a steady state
flow regime has been established and the persistence time distribution
decays with an exponent of $\sim -2.2$.

Figure \ref{fig:creep}(d) shows that the hop displacement distribution
is wider for the $\sigma = 0.5$ case. At this stress, $p(dx)$ depends
on the strain rate (discussed further below), and is highly
anisotropic.
The non-affine displacements in the direction of the applied stress
are significantly smaller than displacements perpendicular to the
stress. This does not appear to be true at any of the smaller
stresses, and is likely an effect of strain hardening during flow. The
fluctuations in the direction of strain may be suppressed in this regime because the
chains are extended and under tension. 

\begin{figure}[tbp] 
\begin{centering} 
\includegraphics[width=8cm]{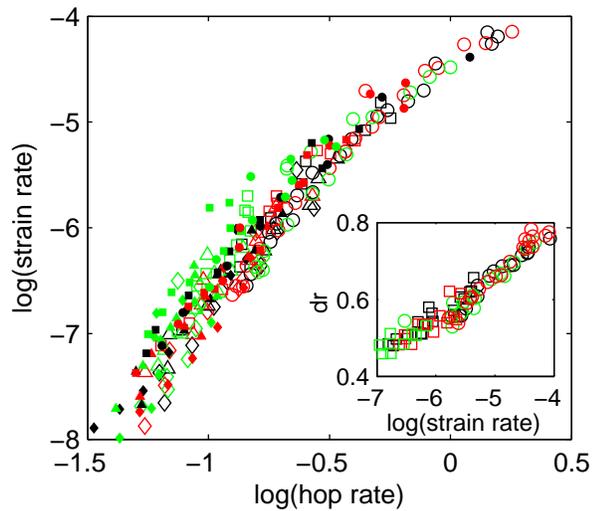}
	\caption{Strain rate versus hop rate for $\sigma = 0.5$
	($\circ$), 0.4 ($\square$), 0.3 ($\triangle$), 0.2
	($\lozenge$) and $t_w = 750$ (black), 7500 (red), and 75000
	(green). Solid symbols show results for chains of length 100.
	Inset shows the mean hop displacement versus the strain rate
	for the same data.  }
	\label{fig:hop_vs_strain_rate} 
\end{centering} 
\end{figure}

Previous experimental \cite{Ediger_JCP128,Ediger_Science323} and
simulation \cite{Ediger_PRL99,Riggleman_MM41} studies have shown a
strong correlation between the particle mobility and the strain
rate. The mobility was defined as the inverse of a relaxation time
$\tau$ obtained from the stretched exponential fits to an autocorrelation function obtained
periodically throughout the creep experiment, and was found to
increase almost linearly with the strain rate. The analogous quantity
to the mobility in our simulations is the average hop rate, which is
determined by creating a histogram of the number of hops as a function
of time. A parametric plot of the strain rate versus the hop rate is
shown in Fig.~\ref{fig:hop_vs_strain_rate}. Data for all stresses and
wait times fall on a universal curve, indicating that there is a
direct relationship between the macroscopic strain rate and the
microscopic hopping transitions.  The strain rate does not increase
linearly with the hop rate, however. Even at zero stress, the hop rate
is finite due to thermally activated transitions. At higher strain
rates, where the hopping dynamics are highly accelerated, the strain
rate increases faster than linearly with the hop rate. This can be
explained by the fact that the average hop displacement also increases
with the strain rate, as shown in the inset. It is possible that this
is due to unresolved hops at very short times, which is discussed
further below.

The polymer chains used to obtain the previous results have a length
of 10 beads, which is far shorter than the entanglement length for
this model. It is important to check how the results depend on chain
length in our polymer model, especially at large strains where strain
hardening is pronounced in entangled polymer glasses. To this end, we
have performed the same simulations with chains of length 100. Figure \ref{fig:creep}(a) (squares) shows that the long-chain glass exhibits considerably more strain hardening during yield, with a lower strain and strain rate than the short-chain glass. We
compare the results of the strain rate versus the hop rate for the two
different chains lengths in Fig.~\ref{fig:hop_vs_strain_rate}. Not
only does the superposition for different wait times and stresses
still hold for the longer chains, but the form of the universal curve
is identical to that found for the shorter chains. 

\subsection{Constant strain rate}
\label{sec:strain_rate}

The stress-strain curve under constant strain rate deformation is
shown in Fig.~\ref{fig:strain_rate}(a). The effects of aging appear as
an increasing over-shoot stress, after which there is a period of
shear softening and eventual flow. The size of this overshoot
increases with the wait time, and with increasing strain rate \cite{Rottler_PRL95}. As in
experiments \cite{Boyce_Polymer34}, the stress on our model glass
in the flow regime is independent of the wait time; all history
dependence has been erased. Because of the low strain rates studied
here, the stress in the flow regime is also independent of the strain
rate.

The probability distributions describing the hop dynamics for this
protocol are shown in Fig.~\ref{fig:strain_rate}(b-d). The constant
strain rate deformation also greatly accelerates the hopping dynamics,
but in a very different way than the step stress. At short times, the
first hop time distributions are markedly unchanged, whereas, at long
times, $p(t_1)$ no longer decays with power law tails but is instead
sharply truncated at a timescale of approximately
$0.1/\dot\epsilon$. The same behaviour is also evident in the
persistence times. In contrast with the case of the step stress,
during constant strain rate flow, $p(\tau)$ does not depend on the
wait time or the measurement time. This is interesting because in the
flow regime strain hardening starts to become important, and yet the
persistence time distribution is the same at short times and at long
times. Conversely, the hop displacements are fairly isotropic at short
times, and become wider and highly anisotropic during the flow
regime. It would appear that strain hardening manifests in the spatial
distribution of hops, rather than their relaxation times.

Mechanical rejuvenation is seen after yield in the constant strain
rate protocol as well, and can similarly be explained by the
persistence time distribution. The truncation in $p(\tau)$ at long times certainly means that the mean
persistence time is finite and a steady state is established in the
flow regime.

\begin{figure*}[tbp]  
\begin{center}
\includegraphics[width=17.5cm]{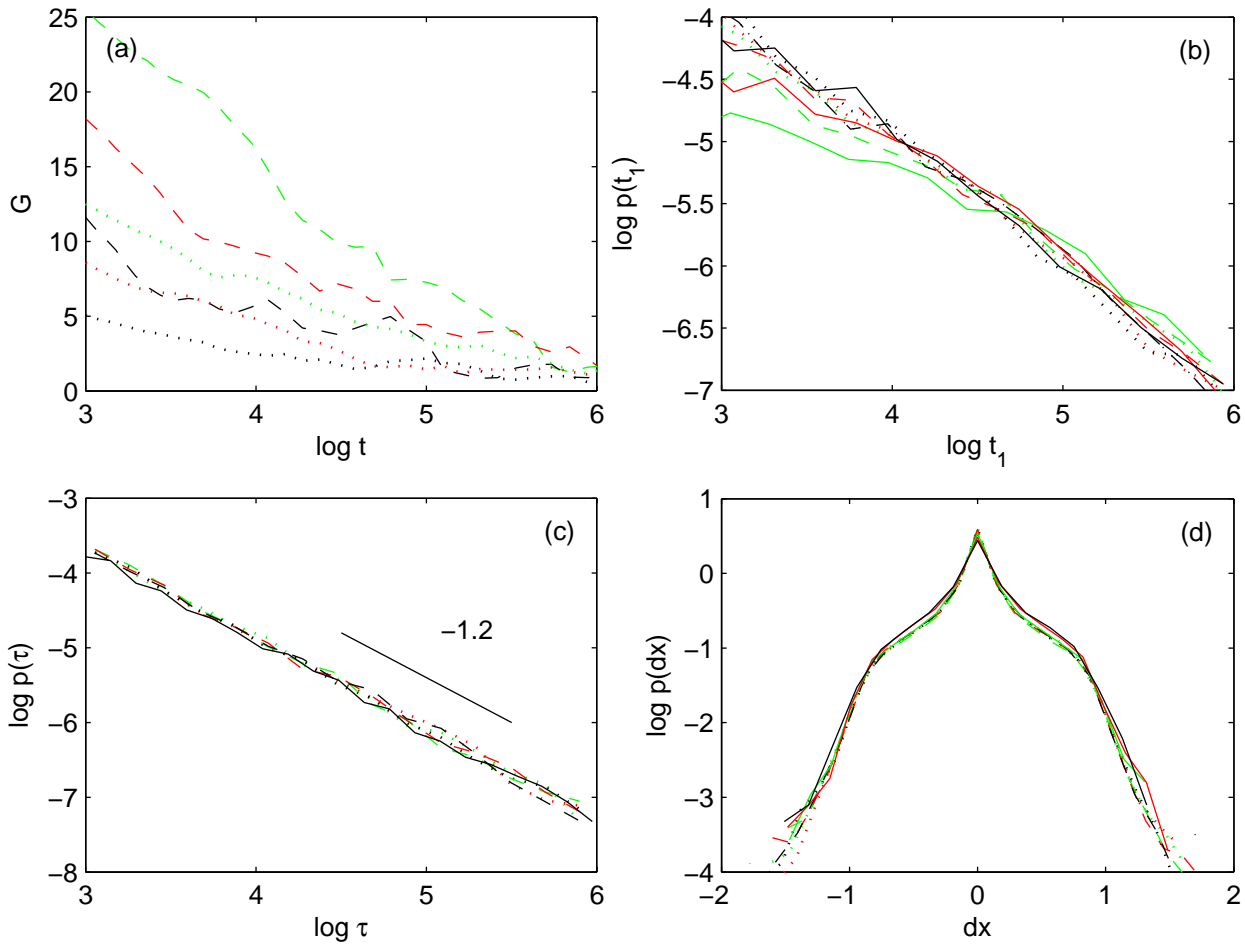}
\caption{Relaxation dynamics during the strain step experiment. (a)
Strain modulus $G = \sigma/\epsilon$ versus time for $\epsilon= 0.02$
(dashed) and 0.04 (dotted). Distributions of (b) first hop times, (c)
persistence times, and (d) displacements for the undeformed sample
(solid lines) and $\epsilon= 0.02$ (dashed) and 0.04 (dotted). In all
plots $t_w = 750$ (black), 7500 (red), and 75000 (green). Straight
lines indicate power law with the given slopes.}
\label{fig:strain}
\end{center} 
\end{figure*}

\subsection{Step strain}
\label{sec:step_strain}

The response to a step strain is defined by the modulus $G(t,t_w) =
\sigma(t,t_w)/\epsilon$ which is shown in
Fig.~\ref{fig:strain}(a). Upon applying the strain, there is an
immediate increase in the stress of a glass, which is larger for
glasses aged for longer wait times. As the strain is held constant,
the stress gradually decays due to structural relaxations over a
timescale which also depends on the wait time. Like the case of the
creep compliance, the step strain modulus is shifted forward in time
with increasing $t_w$. For larger strains outside of the linear
regime, $G(t,t_w)$ decreases with increasing step strain $\epsilon$;
the glass becomes softer because the dynamics are accelerated by the
deformation.

The first hop time distribution (Fig.~\ref{fig:strain}(b)) shows that
the number of relaxations occurring at very short times is greatly
increased by the step strain; however, at longer times, $p(t_1)$
decays with the same power law as the undeformed glass.  At the
largest strains investigated, $p(t_1)$ for all wait times appear to be
almost identical, which would suggest a loss of history dependence
(rejuvenation). The stress relaxation response indicates, however,
that this is not the case. Even at very large strains, the modulus
retains the features of aging. It is likely that the variation in
$p(t_1,t_w)$ is to be found at short times where individual hops
cannot be resolved by our methods. Unlike the case of both the creep
and the constant strain rate experiment, the persistence times
(Fig.~\ref{fig:strain}(c)) and the displacements
(Fig.~\ref{fig:strain}(d)) are completely unaffected by the
deformation: after a first hop, the particle loses all memory of the
strain step. This result implies that the persistence times are not sensitive to
the global stress. There is no time or wait time dependence to the
persistence time distributions, even though the stress relaxation
depends on both times and remains finite for the duration of the
experiment. This is in direct contradiction to the Eyring model of
mechanically accelerated dynamics, which says that the stress
decreases transition barriers and hence increases mobility
\cite{Eyring}.


\subsection{Acceleration Factor}
\label{sec:acceleration}

\begin{figure}[htbp] 
\begin{centering} 
\includegraphics[width=8cm]{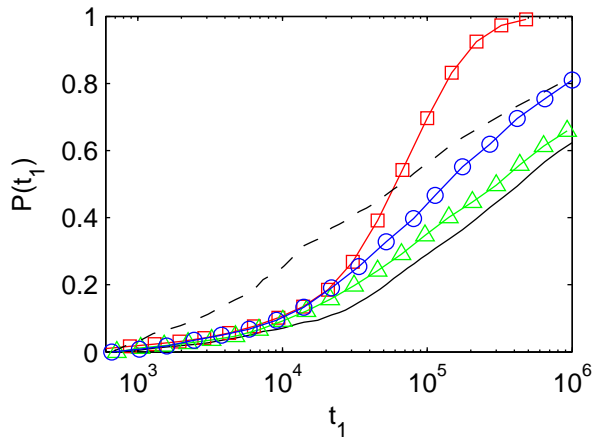}
	\caption{\label{fig:cumulatives} Cumulative probability
          distribution of the first relaxation event for the
          undeformed glass (solid line) and under deformation with a
          step stress $\sigma=0.3$ $(\circ)$, a step strain
          $\epsilon=0.01$ $(\triangle)$, and a constant strain rate
          $\dot\epsilon=8.5\times 10^{-7}$ $(\square)$ for
          $t_w=75000$. Dashed line shows the cumulative for the
          undeformed $t_w = 750$ glass.}
\end{centering} 
\end{figure}

In this section, we develop a method to evaluate the complex
transformations of the relaxation time distributions that were
observed for the three protocols, and to determine which deformation
variables best describe the accelerated dynamics. We assume that the
deformed relaxation times $t_d$ are a function of the undeformed
relaxation times $t_u$ as well as some undetermined parameters
$\vec{p}$ which could include, for instance, the stress, strain,
and/or strain rate: $t_d = t_d(t_u,\vec{p})$. If the
distribution of hop times in the undeformed case is $p_u(t_u)$,
then the distribution in the deformed sample becomes:
\begin{equation}
	p_d(t_d (t_u,\vec{p})) = p_u(t_u)\frac{dt_u}{dt_d}
\label{eqn:diff}
\end{equation}
Integrating on both sides, we obtain
\begin{equation}
\int_0^{t_d}{p_d(t)dt} = \int_0^{t_u}{p_u(t)dt} 
\label{eqn:P}
\end{equation}
where the integrals in Eq.~(\ref{eqn:P}) are just the cumulative
distributions $P_u(t_u)$ and $P_d(t_d)$ of the first hop
times.  The acceleration factor can therefore be defined as
$t_u/t_d$, where $t_u$ and $t_d$ are the times when the
undeformed and deformed cumulative distributions are equal:
$P_d(t_d) = P_u(t_u)$. In this way, we obtain the acceleration
of the entire spectrum of relaxation times during the experiment and
we can determine the form of $t_d(t_u,\vec{p})$, and identify
the relevant parameters $\vec{p}$. This analysis can be performed on both the first hop time and the persistence time distributions.

The cumulatives of the first hop times for representatives of the three deformation protocols are shown in Fig.~\ref{fig:cumulatives}. 
It is clear from this figure that the acceleration factor is unique to
each deformation protocol. The step strain seems to cause a simple
shift in the cumulative distribution with respect to the undeformed
case, which means that all relaxation times are accelerated by a
constant factor. There is also experimental evidence for a simple
rescaling of the relaxation times under a step strain. O'Connell et
al.~found that the stress relaxation after different step strains
could be superimposed by shifting the curves in time, that is, the
glasses obeyed ``strain-time superposition''
\cite{McKenna_O'Connell2002}. Alternatively, in both the stress step and the
constant strain rate experiments, the cumulatives for the deformed
samples continue to diverge from the reference curve at longer
times. This is just another way to see that the distribution of
relaxation times is narrowed: long relaxation times are accelerated
more than short ones. 

\begin{figure}[htbp] 
\begin{centering} 
\includegraphics[width=8cm]{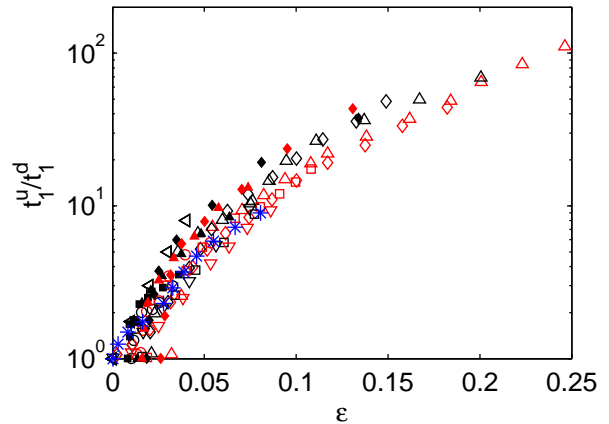}
	\caption{\label{fig:a_vs_strain} Acceleration factor $t_u/t_d$
          as a function of global strain $\epsilon$ for three
          different deformation protocols. Stress step: $\sigma = 0.3$
          ($\circ$), 0.4 ($\square$), 0.5 ($\triangle$); constant
          strain rate: $\dot{\epsilon} = 8.9\times 10^{-6}$
          ($\lozenge$), $8.9 \times 10^{-7}$ ($\triangledown$); strain
          step ($\triangleleft$).  For each: $t_w = 75000$ (black),
          22500 (red). Also shown is the acceleration factor for the
          persistence times for constant strain rate deformation at
          $\dot\epsilon = 8.9 \times 10^{-7}$ (blue $*$). Solid
          symbols show results for chains of length 100.}
\end{centering} 
\end{figure}

We have already seen that the global stress is not a good variable to
describe the dynamical acceleration, and the fact that the step strain
causes a simple shift in $P(t_1)$ suggests the strain as a promising
alternative. In Fig.~\ref{fig:a_vs_strain}, the acceleration factor 
computed from the first hop time distribution $t_1^u/t_1^d$ is
plotted parametrically versus the strain $\epsilon$ for all
protocols. The acceleration factor for the three different protocols
and several wait times collapse quite well onto a common curve. No
similar collapse could be found for the acceleration factor versus any
other macroscopic deformation variable. The acceleration factors computed using representative simulations with chains 100 beads long are also shown in Fig.~\ref{fig:a_vs_strain} and have exactly the same form as the short chain results. 

A strain-controlled acceleration factor also explains the observed
changes to the persistence time distributions. We showed in
Fig.~\ref{fig:traj} that individual particle trajectories exhibit very
little non-affine displacement except during a hop. This suggests that
the local strain tracks with the global strain in between relaxation
events, and that the relevant parameter in analyzing the effect of
deformation on the persistence times is not the total strain since the
onset of deformation, but the accumulated strain in between subsequent
hops. With this interpretation, we understand that the strain step
causes no change to $p(\tau)$ because there is no further strain after
the initial step. Note that this does not imply that there are no residual
effects of the step strain on the dynamics after the first hop has
relaxed the local strain: the persistence time distribution $p(\tau)$
is narrower than the first hop time distribution in an aging glass,
therefore particles displaced by the deformation are more likely to
hop again than those that have not. 
In the constant strain rate experiment, we can compute the acceleration factor of the persistence times $\tau_u/\tau_d$ from the cumulatives of $p(\tau)$ in the same way as before. Plotted versus the global strain increment in between hops (in this case, $\epsilon$ is proportional to the persistence time), Fig.~\ref{fig:a_vs_strain} shows that the acceleration function for the persistence times is identical to that of the first hop times. Although it is more difficult to quantify, the dependence of $p(\tau)$ on the wait time and the
measurement time in the step stress experiment can be explained by the
fact that the strain explicitly depends on these same variables. 

We note here that the acceleration ratios of only the longest wait
time curves could be evaluated with great accuracy. Our method of
identifying hops, whereby the average position and standard deviation
are calculated over a time window, leads to a minimum time resolution
where individual hops can be identified. The relaxation time distributions $p(t_1)$ and $p(\tau)$ are not affected by this resolution limit, as discussed in more detail in the Appendix. However, the fact that $p(t_1)$ and $p(\tau)$ cannot be measured at very short times does presents a problem in
calculating the cumulatives, especially when there are many short time hops, such
as for shorter wait times or very rapid deformations. In this case,
because there are many hops at short times that are not detected, the
cumulative increases abruptly from zero at the beginning of the
measurement, rather than showing the smooth increase typical of the
cage escape regime (see Fig.~\ref{fig:cumulatives}, dashed line). The curve
is effectively shifted downward from where it should be because the
integrated effect of the shortest time hops is not present.
The acceleration factor is thus most accurate at long wait times, where
the cumulative increases slowly from zero, and
where the short time dynamics in particular are not greatly
accelerated, as in the constant strain rate experiment. In the
constant strain rate experiment, the acceleration factor for all wait
times can be evaluated, and the curves are found to collapse. In the
step strain experiment at very large strains, the short time dynamics
are accelerated to such an extent that even the longest wait time
sample may show truncation effects. To obtain a better estimate of the
acceleration factor in this case, we additionally used curve fitting
to find the factor $a$ where $p_d(a\times t_1)/a$ most closely
resembled $p_u(t_1)$ and vice versa, fitting $a\times p_u(t_1/a)$ to
$p_d(t_1)$. The short time resolution of our technique may also be the
cause of the widening displacement distributions with strain rate. If
several hops occur within our minimum resolution, they are detected as
a single hop, possibly with a wider displacement. It is unclear from
our results to what extent the elementary hop displacements actually
grow with the strain rate. 

\begin{figure}[htbp] 
\begin{centering} 
\includegraphics[width=8cm]{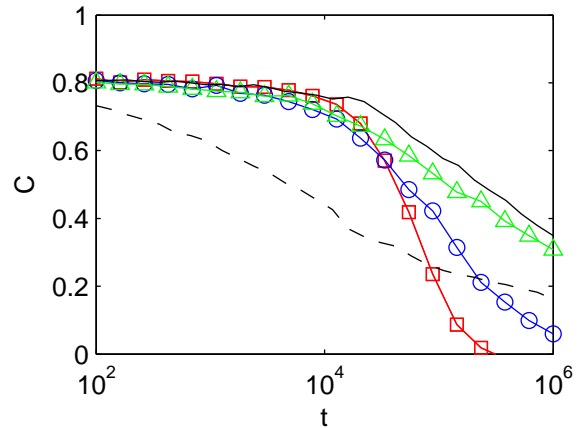}
	\caption{\label{fig:correlation} Incoherent scattering
	        function $C(t,t_w)$ with $|\mathbf{q}| = 7.0$ averaged
	        over the non-deformed ($y$ and $z$) directions for the
	        undeformed glass (solid line) and under deformation
	        with a step stress $\sigma=0.3$ $(\circ)$, a step
	        strain $\epsilon=0.01$ $(\triangle)$, and a constant
	        strain rate $\dot\epsilon=8.5\times 10^{-7}$
	        $(\square)$ for $t_w=75000$. Dashed line shows $C(t,t_w)$
	        for the undeformed $t_w = 750$ glass.}
\end{centering} 
\end{figure}

\begin{figure}[htbp] 
\begin{centering} 
\includegraphics[width=8cm]{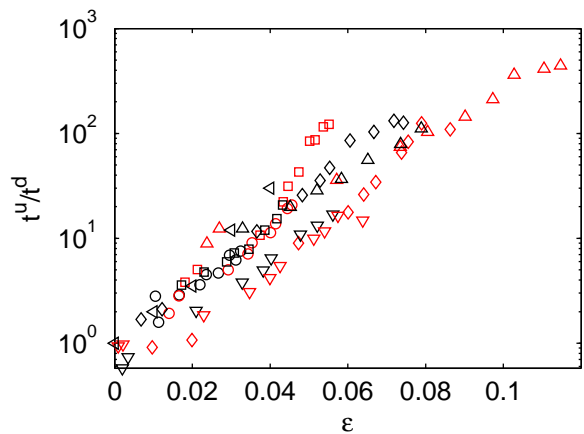}
	\caption{\label{fig:a_from_C} Acceleration factor calculated
					from the incoherent scattering
					function as a function of
					global strain $\epsilon$ for
					three different deformation
					protocols. Stress step:
					$\sigma = 0.3$ ($\circ$), 0.4
					($\square$), 0.5
					($\triangle$); constant strain
					rate: $\dot{\epsilon} =
					8.9\times 10^{-6}$
					($\lozenge$), $8.9 \times
					10^{-7}$ ($\triangledown$);
					strain step ($\triangleleft$).
					For each: $t_w = 75000$
					(black), 7500 (red). }
\end{centering} 
\end{figure}

The hop statistics provide a powerful way of evaluating the full
relaxation time distributions and their history
dependence. Unfortunately, the first hop time and persistence time
distributions are probably not directly accessible in real polymer
glasses. However, it may be possible to calculate the acceleration
factor from the autocorrelation functions more typically used to
measure glassy dynamics. In Fig.~\ref{fig:correlation}, we plot the
intermediate scattering function $C(t,t_w) = \langle \exp[-i \mathbf{q}
\cdot (\mathbf{r}(t,t_w)- \mathbf{r}(0,t_w))]\rangle$ under the same
deformation conditions as the cumulatives in
Fig.~\ref{fig:cumulatives}. Once again the affine portion of the
deformation is removed and, to minimize the effect of possible strain
hardening, $C$ is calculated only in directions perpendicular to the
direction of strain. It can be seen that these curves are
qualitatively similar to $1-P(t_1)$. This close connection is to be
expected: $1-P(t_1)$ measures the number of particles that have not
yet hopped by time $t_1$, which is another measure of the particle
autocorrelation. The intermediate scattering function additionally
includes the effects of multiple hops, since each hop need not be very
large. This effect causes a steeper decay in $C(t,t_w)$ than in $1-P(t_1)$,
especially when the subsequent hops are also accelerated. An
acceleration factor $t^u/t^d$ can be defined for the scattering
function decay in the same way as for the cumulatives through the
relationship $C_u(t^u) = C_d(t^d)$, where $C_u$ and $C_d$ are the undeformed and deformed correlation functions respectively. This acceleration factor is shown
in Fig.~\ref{fig:a_from_C}, and also approximately collapses as a
function of the strain. The curves increase nearly exponentially for all
strains, and more rapidly than the acceleration function calculated from the first hop time distributions. The scatter in this plot is significantly higher than for the acceleration factor calculated from the cumulative distributions, for reasons that are not entirely understood. However, we believe this measure holds promise for
confirming experimentally whether the strain is indeed a good way to
describe deformation-induced acceleration of the relaxation dynamics
in glasses. Our analysis evaluates the full shape change of the particle auto-correlation function due to deformation, rather than reducing the data to an average mobility and width of the correlation decay.

\section{Discussion and Conclusions}
\label{sec:conclusions}

The statistics of single particle hopping dynamics have been evaluated
under three different deformation protocols. All three deformation
modes accelerate the segmental dynamics, although the exact
transformation of the relaxation time distributions is specific to the
mode of deformation. The universal collapse observed in
Fig.~\ref{fig:a_vs_strain} of the acceleration factor as a function of
the strain is strong evidence that the local strain is a good
variable to describe the influence of deformation on structural
relaxations, rather than the global stress as postulated by the Eyring
model. A compelling reason for this is suggested by the behaviour of
the stress and strain at a local level. While the local strains track
very well with the global strain in between relaxation events,
the global stress does not predict the evolution of stress on a
coarse-grained scale \cite{dePablo_PRL93}. The local distributions of
moduli and stress surely account for the wide distribution of
relaxation times \cite{Barrat_PRE80}, however the strain appears to be
a better predictor of how these relaxation times are accelerated.

By tracking the individual particle relaxations, our analysis permits
direct observation of the narrowing of the relaxation time spectrum
during creep and constant strain rate deformation, which previously
was inferred only indirectly from stretched exponential fits to
correlation functions \cite{Ediger_Science323}. The strain-controlled
acceleration factor also provides a simple, intuitive explanation for
the narrowing: particles with long relaxation times experience more
acceleration than particles that hop sooner simply because they
experience more strain in the long period of time in between
relaxations. Alternatively, a step strain leads to a simple shift in
the first hop times (no narrowing), and no change at all to the
persistence times. The fact that the persistence times are unchanged
in the step strain protocol suggests that the strain at the
microscopic level is released after a relaxation event, or reset to
zero. This interpretation is further confirmed by the fact that the
acceleration of the persistence times for the constant strain rate
experiment is identical to that found for the first hop times if,
instead of the total strain, the strain in between subsequent hops is
taken to be the relevant mechanical parameter.

Our results do not imply that the average segmental dynamics in the
glass should be monotonically increased with the strain. In their
experiments measuring the segmental dynamics during creep at stresses
above yield, Lee et al.~found that the average segmental mobility
increases with strain at first, becoming a maximum where the strain
rate was largest, and then decreases again in the strain hardening
regime \cite{Ediger_Science323}. This is completely consistent with
the strain-controlled acceleration factor presented here. The
acceleration factor increases with the strain only in between
subsequent hops; as discussed previously, after a relaxation event the
local ``strain clock'' is reset to zero. As a result, at the beginning
of a deformation experiment, the local strains track closely with the
global strain; however, after some time, the local strains in the
glass become spatially heterogeneous, being higher in regions that
have not yet relaxed, and lower in regions that have recently
experienced a relaxation. In the steady state flow regime, our
simulations show that every particle has hopped at least once and
usually many times. The advantage of molecular dynamics simulations is
that the individual trajectories reveal exactly how much time has
elapsed between local relaxations and thus how much strain has
accumulated in each local region.

It is interesting to note that the strain was chosen as control
variable in the Soft Glassy Rheology (SGR) model
\cite{Sollich_JR44}. Here the acceleration factor for barrier
crossings increases proportional to $\exp[kl^2/2x]$, where $k$ is the
elastic constant of a mesoscopic element, $l$ a measure of local
strain on that element and $x$ denotes a noise temperature. The local
strains increase in tandem with the global strain in between barrier
crossings, and after a hop, the mesoscopic element returns to a
position of zero local strain. The main elements of the SGR model
closely resemble our findings for the single particle relaxation
times. The measured form of the acceleration factor is, however, quite
different. It would be interesting to see if using the measured form
of the acceleration factor could account for some of the limitations
identified in this model \cite{Warren_PRE78}.

The hop statistics under deformation also present an explanation for
mechanical rejuvenation. In contrast to the case of sub-yield
deformations where the response continues to show explicit history
dependence, the effects of aging are lost once the glass has
yielded. Non-stationary dynamics occurs in glasses because of the wide
distribution of persistence times. An infinite $\langle \tau \rangle$
means that an equilibrium state is impossible, and the dynamics experiences aging. During yield in
both the step stress and the constant strain rate protocols, the
persistence time distributions are significantly narrowed, and
stationary dynamics are restored. The resulting rejuvenated state is,
however, not the same as a just-quenched glass. The relaxation time distributions depend explicitly on the
deformation protocol. Further investigation of the relaxation time
distributions after deformation may shed more light on the nature of
the rejuvenated state.

The hop dynamics under deformation also provide some information about strain hardening in polymer glasses. We have tested two different chain lengths and find that the hop rate versus strain rate and the acceleration versus strain are identical for both chain lengths. Even though strain hardening is significantly more pronounced in the long-chain glass, the deformation appears to affect the hop dynamics in the same way. Strain hardening is most evident in the displacement distribution, which becomes anisotropic and significantly narrower in the strain direction. Further study of the hop dynamics during strain hardening may provide further insight into this very important behaviour of polymer glasses.

In conclusion, by detecting individual segmental relaxations, the full
transformation of the relaxation time distributions under deformation
could be quantified and correlated with the global deformation
parameters in a model polymer glass. We find that the accelerated
segmental dynamics are fully explained by the global strain, and we
present a possible experimental test of these findings though the
definition of the acceleration factor for any segmental
autocorrelation function such as the incoherent scattering function.

\begin{acknowledgments}
We thank M.~D.~Ediger for many helpful discussions. We acknowledge the Natural Sciences and Engineering Council of Canada
(NSERC) for financial support. Computing time was provided by
WestGrid. Simulations were performed with the LAMMPS molecular
dynamics package \cite{LAMMPS}.
\end{acknowledgments}

\appendix*
\section{Identifying hopping events}

\begin{figure}[htbp]  
\begin{center}
\includegraphics[width=8cm]{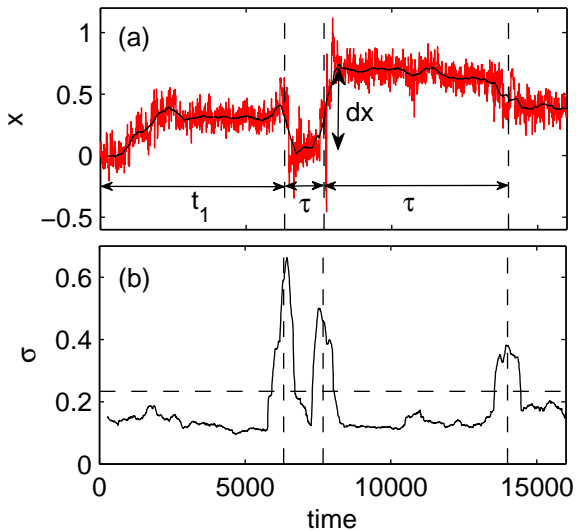} 
\caption{(a) A typical particle trajectory in the glass. Red curve shows raw position data in the x-direction, and the black curve shows the running average. (b) Standard deviation in the three dimensional particle position over the averaging window. Hops are identified by a threshold in the standard deviation, shown here as a horizontal dashed line, and marked in both frames as vertical dashed lines. In panel (a) the first hop time, persistence times, and particle displacements are also labeled.}
\label{fig:traj_app}
\end{center} 
\end{figure}

\begin{figure}[htbp]  
\begin{center}
\includegraphics[width=8cm]{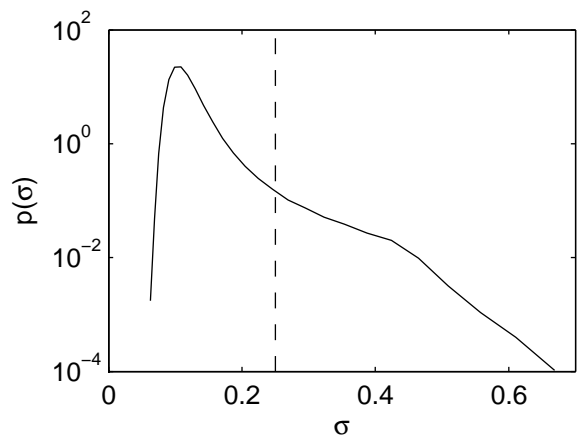} 
\caption{Probability distribution of the standard deviation
$\sigma$ in the particle position over an averaging time window of 400
for all particle trajectories. Dashed line shows the threshold
$\sigma_{th}$ for detecting a hop. }
\label{fig:p_sigma}
\end{center} 
\end{figure}

\begin{figure*}[tbp]  
\begin{center}
\includegraphics[width=15cm]{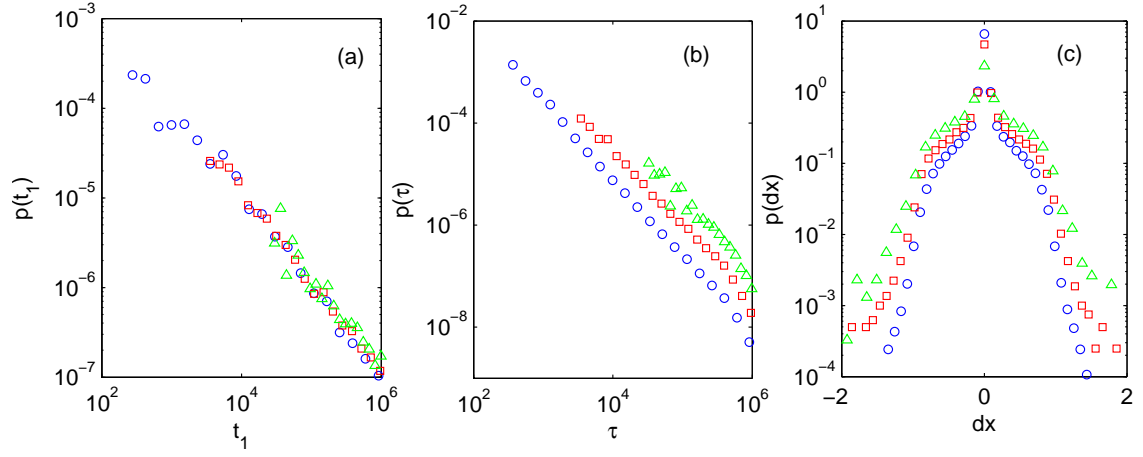}
\caption{The distributions (a) $p(t_1)$, (b) $p(\tau)$ and (c) $p(dx)$
for $t_{dump} = 3.75$ (circle), 37.5 (square) and 375 (triangle). }
\label{fig:T_dump}
\end{center} 
\end{figure*}

\begin{figure*}[tbp]  
\begin{center}
\includegraphics[width=15cm]{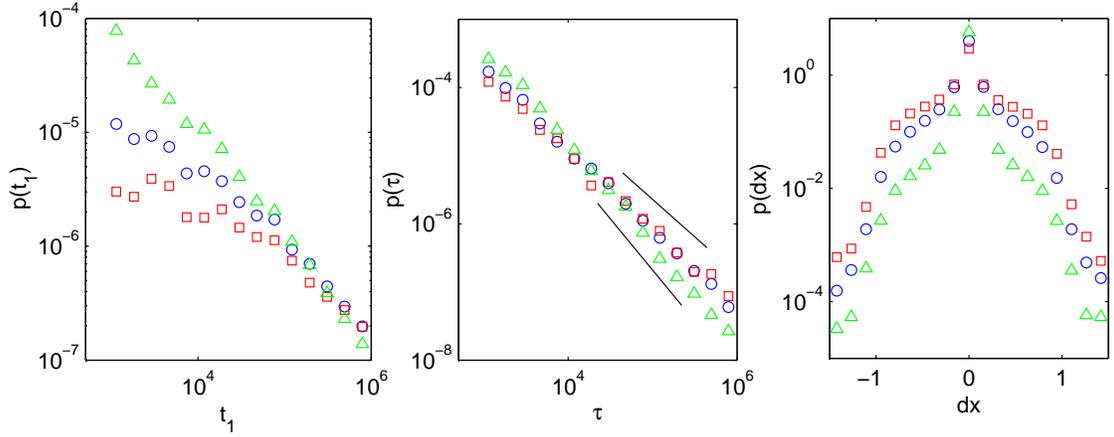}
\caption[Hop statistics with different values of the standard
deviation threshold]{The distributions (a) $p(t_1)$, (b) $p(\tau)$ and
(c) $p(dx)$ for $\sigma_{th} = 0.15$ (square), 0.25 (circle), and 0.35
(triangle). Solid lines in (c) indicate power laws with slope of -1.5 and -1.1.}
\label{fig:thresh}
\end{center} 
\end{figure*}

\begin{figure*}[tbp]  
\begin{center}
\includegraphics[width=15cm]{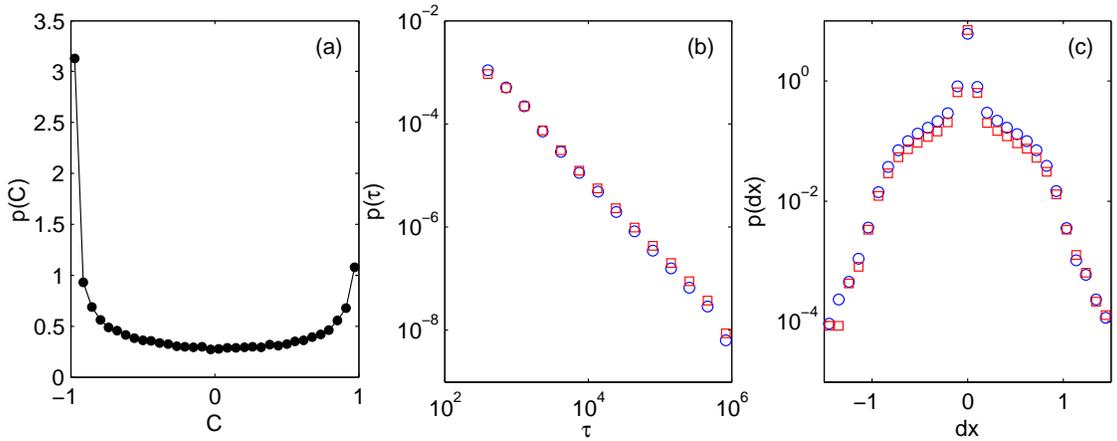}
\caption{(a) The distribution of hop correlations, (b) $p(\tau)$, and (c) $p(dx)$ computed with all of the hops detected (circles), and with only uncorrelated hops (squares).}
\label{fig:corr}
\end{center} 
\end{figure*}

Particle positions are written to file with a period of
$t_{dump}$. The trajectories show rapid vibrations about a mean
position, punctuated by the occasional hop to a new location. To
detect the hopping transitions, the average and standard deviation of
each particle position is computed over $N_{avg}$ snapshots at times
$t_i = N_{avg}\times t_{dump}\times(i-1/2)$. The trajectories and standard deviation are shown in Fig.~\ref{fig:traj_app}. The value of $N_{avg}$ used
for determining the mean and standard deviation is on the order of
$40-60$ snapshots for the work presented here. Increasing $N_{avg}$ by
a factor of 10 (from 40 to 400), causes negligible changes to the hop
statistics but seriously decreases the time resolution of the
distributions. The criteria for finding a hop is that the standard
deviation of the particle position over $N_{avg}$ snapshots is greater
than a threshold: $\sigma>\sigma_{th}$. To choose a suitable
threshold, we look at the distribution of the standard deviations over
the full time series of all of the particles in the system (see
Fig.~\ref{fig:p_sigma}). There are clearly two regimes of motion in
$p(\sigma)$: a narrow vibrational regime and a hopping
regime. $\sigma_{th}$ is chosen to be at the shoulder in between these
regimes and is shown in Fig.~\ref{fig:p_sigma} as a dashed line. The effects of the threshold on the distributions is discussed below.

The times at which a hop occurs are determined by the centre of the
standard deviation peak, and the displacements are found from the
positions on either side of the peak. The distribution of displacements is easily computed
by binning the data to form a histogram. Because the relaxation time
distributions are very broad, the bin sizes for $p(t_1)$ and $p(\tau)$ increase logarithmically
with time in order to obtain good statistics. The finite timespan of
the simulation means that this distribution is necessarily
truncated. In other words, some of the particles will not have hopped
at all during the experiment. For the first hop times $t_1$, this can
be accounted for by normalizing to the total number of particles in
the ensemble, rather than the number of first hops detected. The
persistence time $\tau$ is defined as the time in between all
subsequent hops. Accounting for the effects of truncation on this
distribution is somewhat more complex than in the case of $p(t_1)$. If
all of the hops are assembled into a single distribution function as
described above, there is a sharp downturn in $p(\tau)$ near the
maximum time observed $t_{max}$. The distribution of persistence times
that we observe $p_{obs}(\tau)$ is weighted by the probability that we
will observe it within $t<t_{max}$, which becomes low for cages originating near the end of the simulation time. This effect can be accounted for using multiple convolutions of the relaxation time distributions; however since the downturn is very sharp, it is in practice easier to simply discard the very long time data.

The effect of the dump frequency on the distributions was investigated
by recording the same data at three different values of
$t_{dump}$. Figure \ref{fig:T_dump} shows the effect of $t_{dump}$ on
(a) $p(t_1)$, (b) $p(\tau)$ and (c) $p(dx)$. Changing $t_{dump}$ does
not affect the first hop time distribution $p(t_1)$, aside from the
obvious relationship between $t_{dump}$ and the minimum $t_1$ that can
be observed. Distributions recorded at different values of $t_{dump}$
are in perfect agreement where their time spans overlap.  The
persistence time distribution $p(\tau)$ has the same slope for all
values of $t_{dump}$, however the curves differ by a normalization
factor which depends on the minimum time resolution in $\tau$. An
absolute normalization factor may be obtained by using a similar
method to that used to normalize the distributions $p(t_1)$: the time
between the first and second hops are found for a subset of $N$
particles where $t_1$ is very short. Not all of these particles will
experience a second hop, and if the distribution is normalized to $N$
rather than the number of second hops, a normalized distribution is
obtained that does not depend on $t_{dump}$. This method requires
throwing away most of the persistence time data, however, and is
therefore only used to properly normalize the full distribution which
includes all measured persistence times. The displacement distribution
$p(dx)$ becomes somewhat wider with increased $t_{dump}$ due to the
effects of undetected hops on timescales shorter than $t_{dump}$. The
distributions compared here have identical $t_{dump}$.

The effect of the threshold on the standard deviation was also
investigated using three different values: $\sigma_{th} = 0.15$, 0.25,
and 0.35. The distributions computed using
these thresholds are compared in Figure \ref{fig:thresh}. Changing the
threshold results in quantitative, rather than qualitative changes to
the distributions. In all cases, the hop time distributions take the
form of power laws, although the exponent is modified by the threshold
(from about -1.1 to -1.5 for $p(\tau)$), and the first hop
time distribution depends on $t_w$ whereas the persistence time
distribution does not. However, if the threshold is very small, the
influence of wait time is decreased because of noise (vibrations
rather than hops) that does not depend on $t_w$; if $\sigma_{th}$ is
very large, the statistics become poor and many legitimate hops are
discarded. The hop displacements predictably become larger with
increasing threshold. Using a continuous time random walk
\cite{Warren_EPL88}, we find that the mean squared displacement and van
Hove function are self-consistently described by the hop statistics
obtained using a range of thresholds near the shoulder of $p(\sigma)$
(Fig.~\ref{fig:p_sigma}).

In order to model the trajectories as a continuous time random walk we assume that the hops are completely uncorrelated. To ensure that this is a reasonable approximation, the correlations between adjacent hops in the trajectories are evaluated. A correlation function can be defined for each pair of displacement vectors $\mathbf{r}_i$
\begin{equation}
	C_i = \frac{\mathbf{r}_{i+1}\cdot \mathbf{r}_i}{\left\| \mathbf{r}_{i+1} \right\| \left\| \mathbf{r}_i \right\|}.
\end{equation}
The distribution of these correlations are evaluated for all pairs of hops, as shown in Fig.~\ref{fig:corr}(a). A completely uncorrelated random walk would have a flat distribution; however, we see that there are a number of backward correlated ($C=-1$) and forward correlated ($C=1$) hops. If these hops are removed from the time series data and the distribution of persistence times and displacements are computed with only the uncorrelated hops, we see no appreciable difference in the hop statistics (Fig.~\ref{fig:corr}(b) and (c)).

\end{document}